# Speech Enhancement Modeling Towards Robust Speech Recognition System

Ms. Urmila N. Shrawankar
G H Raisoni College of Engg,
RTM Nagpur University
NAGPUR-INDIA
Cell : +91-9422803996

urmilas@rediffmail.com

Dr. V. M. Thakare
SGB Amravati University
Amravati-INDIA

## ABSTRACT
For about four decades human beings have been dreaming of an *intelligent machine* which can master the natural speech. In its simplest form, this machine should consist of two subsystems, namely automatic speech recognition (ASR) and speech understanding (SU). The goal of ASR is to transcribe natural speech while SU is to understand the meaning of the transcription. Recognizing and understanding a spoken sentence is obviously a knowledge-intensive process, which must take into account all variable information about the speech communication process, from acoustics to semantics and pragmatics.

While developing an Automatic Speech Recognition System, it is observed that some adverse conditions degrade the performance of the Speech Recognition System.

In this contribution, speech enhancement system is introduced for enhancing speech signals corrupted by additive noise and improving the performance of Automatic Speech Recognizers in noisy conditions.

Automatic speech recognition experiments show that replacing noisy speech signals by the corresponding enhanced speech signals leads to an improvement in the recognition accuracies. The amount of improvement varies with the type of the corrupting noise.

## Categories and Subject Descriptors
### Speech User Interface (SUI) for Human Computer Interaction (HCI):
**H**uman-Computer-Interaction is not simple as human-to-human interaction. Human-to-human interaction mainly based on speech, emotion and gesture, where as Human-Computer-Interaction is based on either text interface or Graphical interface (GUI).
If we provide an artificial intelligence to train a machine such a way, so that machine may interact with speech signals.

## 1 INTRODUCTION

Speech is a fundamental means of human communication. In the last several decades, much effort has been devoted to the efficient transmission and storage of speech signals. With advances in technology making speech user interface (SUI) successful. The freedom and flexibility offered by SUI brings with it new challenges, one of which is robustness to acoustic background noise. Speech enhancement systems form a vital front-end for speech operated applications in noisy environments such as in offices, cafeterias, railway stations, etc., to improve the performance of speech recognition systems.

While these technologies show impressive performance in controlled noise-free environments, performance rapidly degrades under practical noisy conditions. Noise reduction is also becoming an increasingly important feature. For these reasons, much effort has been devoted over the last few decades towards developing efficient speech enhancement algorithms. The term speech enhancement has a broad connotation encompassing various topics such as acoustic background noise reduction, dereverberation, blind source separation of speech signals, bandwidth extension of narrowband speech, etc.

This literature review paper explains a framework to exploit available prior knowledge about both speech and noise. The physiology of speech production places a constraint on the possible shapes of the speech spectral envelope, and this information is captured using codebooks of speech linear predictive (LP) coefficients obtained from a large training database. Similarly, information about commonly occurring noise types is captured using a set of noise codebooks, which can be combined with sound environment classification to treat different environments differently.

The speech pdf may be described using a Laplacian density and the noise pdf using a Gaussian density. A more accurate method, though computationally more demanding, is to use more sophisticated statistical models using, e.g., hidden Markov models (HMMs), Gaussian mixture models (GMMs), or codebooks that have been trained using a representative database. The pdfs of the speech and noise processes are thus estimated from corresponding training sequences. The adapted approach uses where prior knowledge about the speech and noise signals, in the form of trained codebooks of their LP coefficients, is used in the estimation procedure. If the parameter is assumed to be deterministic (but unknown), the procedure is termed classical estimation, e.g., maximum-likelihood (ML) estimation. If we assume that the unknown parameter is a random variable with its own pdf, and we estimate a realization of that random variable, the procedure is termed Bayesian estimation.

## 2. NOISE ESTIMATION AND SPEECH ENHANCEMENT MODELS

### 2.1 Wiener Filtering:

In the Wiener filter approach, the optimal estimator is designed to minimize the mean squared error.

### 2.2 Spectral Subtraction:

Spectral subtraction is a speech enhancement scheme based on a direct estimation of the short-time spectral magnitude of clean speech. The estimated magnitude is combined with the noisy phase.

### 2.3 Subspace Based Models:

The Wiener filter does not make a distinction between the two types of distortions. An alternate approach motivated by perceptual considerations is to have a trade-off between noise reduction and signal distortion, and was introduced as subspace methods.

### 2.4 Kalman Filter

Wiener filtering, spectral subtraction and the subspace methods discussed above can generally be categorized as non-parametric methods in the sense that they do not employ any parametric model to describe the speech signal.

This is in contrast with parametric methods that use models such as the Autoregressive (AR) or the Sinusoidal model to describe the signal. We discuss here one specific approach, the Kalman filter, which provides a framework that can exploit information about the human speech production process by using the AR model.

### 2.5 Statistical Models

In the above mentioned models, we considered linear estimation techniques for the signal. Linear estimation is optimal (in the Mean squared error-MSE sense) for the case when x and y are jointly Gaussian. The Wiener filter represents the optimal solution in this case. In this section, we look at methods that use distributions other than Gaussian and derive optimal nonlinear solutions.

We first consider methods that retain the Gaussian assumption on the speech and noise processes in the frequency domain, i.e., the respective Discrete Fourier Transform - DFT coefficients are assumed to be normally distributed. They differ from the Wiener solution in that they attempt to obtain MMSE estimates of the spectral amplitude, which then follows a Rayleigh distribution. Here, methods are also discussed that assume super-Gaussian (Gamma, Laplace etc.) models.

*2.5.1 Gaussian Models*

In the Wiener filter approach to speech enhancement, an optimal (under the Gaussian assumption and in the mean squared error sense) estimate of the clean speech spectral component is obtained from the noisy speech. The spectral amplitude is perceptually more relevant than the phase and thus performance could be improved by an optimal estimate of the amplitude. The amplitude estimate provided by the Wiener filter (obtained as the modulus of the optimally estimated spectral component) is not optimal under the assumed model; only the estimate of the spectral component is optimal. Using the same statistical model, an optimal estimate of the spectral amplitude, with noisy speech can be obtained.

The Fourier expansion coefficients of the speech and noise processes are assumed to be independent zero mean Gaussian variables with time-varying variances. This results in a Rayleigh distribution for the amplitudes of the Fourier coefficients.

The MMSE estimate of the complex exponential of the phase can be derived and used together with the MMSE amplitude estimate. It is shown that the modulus of the resulting estimate of the phase is not unity. Thus combining the MMSE phase estimate with the MMSE amplitude estimate affects the optimality of the amplitude. To address this problem, a constrained MMSE estimate of the phase is obtained, whose modulus is constrained to be unity. The resulting constrained MMSE estimate is the noisy phase itself. The MMSE amplitude estimate is obtained by applying gain function to the noisy spectral magnitude.

*2.5.2 Super-Gaussian Models*

The methods discussed in the previous section assume that the speech DFT coefficients follow a Gaussian distribution. In this section, we discuss methods that assume a super-Gaussian distribution. Super-Gaussian random variables, also called leptokurtic, have a positive kurtosis. They have a more peaky pdf (Probability density function) than Gaussian random variables and possess heavier tails, e.g., Laplace and Gamma distributions. The DFT coefficients of speech are better modeled by a Gamma distribution. Under a Gaussian assumption for speech and noise, the estimator is linear (Wiener filter). Assuming a Gamma distribution for speech and either a Gaussian or Laplacian distribution for noise, two non-linear MMSE estimators of the complex DFT coefficients are derived. Experimental results show a small but consistent improvement in terms of SNR-Signal-to-noise ratio over the Wiener filter. For high a-priori SNR (e.g., 15 dB) the estimator exhibits a behavior similar to the Wiener filter.

For the case when a Laplacian model is used for noise, for low a-priori SNR (e.g., -10 dB), the attenuation is constant regardless of the magnitude of the noisy DFT coefficient, resulting in reduced musical noise.

Assuming a Gamma distribution for speech and Laplace or Gaussian for noise, MMSE estimates of the squared magnitude of the speech DFT coefficients can be obtained. It was observed that using a Gaussian model for the noise signal resulted in musical noise, which was avoided by the Laplace model.

MMSE estimation of the complex DFT coefficients under a Laplacian model for speech is discussed in referenced papers. The resulting estimator has a simpler analytic form compared to the case when a Gamma prior was used for speech. Using a Laplace model for noise as well results in less musical noise. MMSE and ML- Maximum-likelihood estimates are obtained in the DCT- Discrete cosine transform domain using a Laplace model for speech and a Gaussian model for noise.

A Gaussian noise model and a super-Gaussian speech model were found to provide a higher segmental SNR than the Wiener filter, which assumes a Gaussian model for both speech and noise. Using a Laplacian noise model was found to achieve better segmental SNR only for high input-SNR conditions, but resulted in more natural residual noise. The Laplacian speech model was favored over the Gamma model as it resulted in lower musical noise. In comparison to the Ephraim-Malah amplitude estimators, the super-Gaussian schemes achieve a higher segmental SNR but the residual noise was found to be less natural. Adaptive a-priori SNR smoothing and limiting are suggested for improving the quality.

## 2.6 Trained Statistical Models

### 2.6.1 HMM Based Methods

HMMs have been used extensively in speech recognition. HMMs trained on clean speech and noise was used for speech enhancement, and Bayesian MMSE and MAP-Maximum a-posteriori estimates of the speech signal were obtained. The HMMs consist of several states with a mixture of Gaussian pdfs at each state. A state transition matrix governs the transition from one state to another. The covariance matrix of each Gaussian pdf is parameterized by the AR parameters of the signal. The AR parameters are the linear predictive coefficients and the variance of the excitation signal.

In the MAP approach, an estimate of the speech signal is obtained by maximizing the posterior pdf of the speech signal given the noisy observations. Since the corresponding gradient equations are nonlinear, a local maximization is performed using the Expectation maximization (EM) algorithm. In the MMSE approach, a weight is associated with the Wiener filter corresponding to each combination of speech and noise components at each state. The MMSE estimate of the clean speech signal is obtained by filtering the noisy signal with the weighted sum of these Wiener filters over all combinations of states and mixtures.

As mentioned earlier, the HMM models both the Linear prediction-LP coefficients and the excitation variance (gain). This generally leads to a mismatch in the gain term between training and testing. Thus some form of gain adaptation is essential. For the MAP estimation, gain-normalized HMMs are trained for the clean speech signal.

### 2.6.2 Codebook Based Approach

The codebook based approaches attempt to overcome the disadvantage of the Hidden Markov model (HMM) methods in non-stationary noise. An instantaneous frame-by-frame gain computation approach was also considered in a speech decomposition context. Using trained codebooks of only the LP coefficients of speech and noise, the gain terms are computed for each short-time frame based on the LP coefficients and the noisy observation. The codebooks are trained using representative databases of speech and noise.

## 2.7 Comparison between HMM and Codebook Approaches

The main difference between the HMM methods and the codebook approaches lies in the manner in which they handle the non-stationary of the noise signal, which in turn is related to the modeling and computation of the excitation variances.

Since the HMM method models both the LP coefficients and the excitation variance as prior information, a gain adaptation is required for the speech and noise models to compensate for differences in the level of the excitation variance between training and operation. The gain adaptation factor is computed using the observed noisy gain and an estimate of the noise statistics obtained using, e.g., the minimum statistics approach. Conventional noise estimation techniques are buffer-based techniques, where an estimate is obtained based on a buffer of several past frames. Thus, such a scheme cannot react quickly to non-stationary noise. In the codebook based approach, the codebook models only the LP coefficients and the speech and noise excitation variances are optimally computed on a frame-by-frame basis, using the noisy observation. This enables the method to react quickly to non-stationary noise. The frame-by-frame gain computation is based on codebook methods, an enhancement scheme with explicit noise gain modeling and on-line estimation are based on HMM.

Another difference is that the HMM based methods obtain MMSE estimates of the clean speech signal whereas the codebook approach obtains MMSE estimates of the speech and noise Short-term predictor (STP) parameters. Let the vector X denote the random variable corresponding to a frame of the clean speech signal. Given the noisy observations, the HMM method obtains the expected value of X and its functions such as the spectral magnitude and the log-spectral magnitude. The codebook method obtains the expected value of μ given the noisy observations for the current and previous frames, which is useful in applications that require optimal estimates of the speech and noise AR parameters. The framework developed in the codebook approach also allows the MMSE estimation of functions of the speech and noise AR parameters, where the MMSE estimate of one such function can be shown to result in the expected value of X given the noisy observations, which is useful in applications where an optimal estimate of the time domain speech waveform is desired.

## 3. CONCLUSION

This literature survey based paper discusses with the enhancement of speech signals that have been subject to acoustic background noise. An estimation-theoretic approach to exploit prior knowledge about the speech and noise signals is developed using maximum-likelihood and Bayesian MMSE estimation. The use of prior information is shown to result in good performance in practical environments with non-stationary background noise.

As discussed earlier, the HMM based methods and the codebook based approaches employ a more accurate model for the speech pdf compared to the methods of Wiener filter and Statistical model. The price to be paid for the improved accuracy is an increase in computational complexity. The complexity is directly related to the model size, e.g., the number of codebook vectors, or the number of states and mixture components in the HMM. The HMM and codebook approaches lend themselves in a straightforward fashion to parallel processing, which can result in a significant speedup.

The amount of time required for the resulting computations is independent of the model size. An additional step of weighted summation is required for the MMSE approaches, though the computation of the likelihood can still be performed in parallel.

## Abbreviations

AR - Autoregressive

DFT - Discrete Fourier transform

MMSE - Minimum mean squared error

MSE - Mean squared error

pdf - Probability density function

SNR - Signal-to-noise ratio

ML - Maximum-likelihood

DCT - Discrete cosine transform

MAP - Maximum a-posteriori

EM - Expectation maximization

LP - Linear prediction

HMM - Hidden Markov model

STP - Short-term predictor